\documentclass[
	a4paper, % Paper size, use either a4paper or letterpaper
	10pt, % Default font size, can also use 11pt or 12pt, although this is not recommended
	twoside, % Two side traditional mode where headers and footers change between odd and even pages, comment this option to make them fixed
]{LTJournalArticle}

\usepackage{comment}
\usepackage{amsmath}
\usepackage{amssymb}

\usepackage{tikz}
\tikzset{every node/.style={font=\small}}
\usepackage{pgfplots}
\pgfplotsset{compat=1.18}
\usepgfplotslibrary{groupplots}

\usepackage{caption}

\usepackage{hyperref}
\usepackage[noabbrev,nameinlink]{cleveref}

\title{A Critical Assessment of the Sample-Based Quantum Diagonalization for Heisenberg and Hubbard Models} % Article title, use manual lines breaks (\\) to beautify the layout

\author{%
Cedric Gaberle\textsuperscript{1}{\thanks{Corresponding author: \href{mailto:gaberle@uni-frankfurt.de}{gaberle@uni-frankfurt.de}}} and Manpreet S. Jattana\textsuperscript{1}\\ 
}

\date{\footnotesize\textsuperscript{\textbf{1}}Modular Supercomputing and Quantum Computing, Institute of Computer Science, Goethe University Frankfurt, 60325 Frankfurt am Main, Germany\\ }

\renewcommand{\maketitlehookd}{%
	\begin{abstract}
        \noindent Sample-based quantum diagonalization (SQD) constructs subspaces 
from computational-basis configurations obtained via measurements of a quantum 
state, with the goal of approximating low-energy eigenspaces of many-body 
Hamiltonians. The effectiveness of this approach relies on the assumption that 
physically relevant states admit a compact representation in the computational 
basis. We investigate this assumption by analyzing SQD subspaces 
constructed directly from configurations of exact ground states of Heisenberg 
and Hubbard model lattices. By eliminating state-preparation and measurement 
inefficiencies, we isolate the intrinsic configuration-space structure of the 
wavefunction. We determine the minimal number of configurations required to 
reproduce the ground-state energy within fixed accuracy thresholds and find that this 
number grows exponentially with the system size. Notably, this scaling persists 
even under optimal inclusion of configurations in order of decreasing probability, 
demonstrating that it originates from intrinsic delocalization of the 
wavefunction rather than sampling inefficiencies. Our results indicate that 
SQD effectively probes the configuration-space entropy but faces 
fundamental scalability limitations for these models.% as the required
	\end{abstract}
}

\begin{document}

\maketitle % Output the title section

%----------------------------------------------------------------------------------------
%	ARTICLE CONTENTS
%----------------------------------------------------------------------------------------

%------------------------------------------------------------------------
\section{\label{sec:Introduction}Introduction}%
%------------------------------------------------------------------------

%Subspace quantum diagonalization relies on the assumption that the low-energy eigenspace of a Hamiltonian can be represented within a manageable configuration subspace generated from a physically motivated reference state. To isolate intrinsic limits of this assumption, we analyze the Sample-based Quantum Diagonalization (SQD) method constructed solely from computational-basis configurations of the exact ground state for lattice Hamiltonians. 
%
%Subspace quantum diagonalization techniques, such as SQD, focus on projecting a Hamiltonian onto a smaller subspace, which simplifies classical eigendecomposition calculations. Instead of diagonalizing the entire Hamiltonian, only the effective subspace that is relevant for the problem at hand is processed classically. 

Determining low-energy eigenstates of interacting quantum many-body Hamiltonians is a central problem in condensed matter physics~\cite{electronic_structure_study,many_particle_physics}, quantum chemistry~\cite{modern_quantum_chemistry,quantum_chemistry,algorithms_for_quantum_chemistry_and_materials_science}, and quantum simulation~\cite{quantum_simulation}. Exact diagonalization methods provide numerically exact solutions but scale exponentially with the system size, limiting their applicability to relatively small systems~\cite{exact_diagonalization,exact_diagonalization_techniques,Lanczos}. Hybrid quantum–classical algorithms have, therefore, been proposed as a promising approach to extend the range of tractable problems by leveraging quantum devices to represent many-body states and interactions while retaining classical post-processing. 

A prominent class of such approaches is quantum subspace methods, which approximate the low-energy eigenspace of a Hamiltonian by projecting it onto a reduced subspace~\cite{quantum_subspace_methods}. Rather than diagonalizing the full Hamiltonian in the exponentially large Hilbert space, these methods construct a set of basis states that are expected to capture the relevant physics and perform classical diagonalization within the resulting subspace. Variants of this strategy appear in quantum subspace expansion methods~\cite{quantum_subspace,partitioned_QSE,qse_greens_functiones}, Krylov-based approaches~\cite{krylov_subspace,quantum_filter_diagonalization,krylov_large_hamiltonians}, and measurement-based subspace algorithms~\cite{SQD,SKQD,symmetry_adapted_SQD,SKQD_Heisenberg}.

Sample-Based Quantum Diagonalization (SQD) is a recently proposed realization of this idea in which the subspace basis is generated directly from measurement outcomes of a quantum state~\cite{SQD_original}. In this framework, repeated measurements of a reference state in the computational basis produce configurations according to their underlying probability. The observed configurations are used to construct a subspace, and the Hamiltonian projected onto this subspace is subsequently diagonalized classically. The effectiveness of this approach relies on the assumption that the dominant configurations of the reference state provide a compact representation of the relevant low-energy eigenspace.

Despite the growing interest in SQD and related methods, the fundamental compressibility of many-body wavefunctions in the computational basis remains poorly understood. %In particular, it is unclear whether the configuration support of physically relevant ground states admits a polynomially bounded representation when configurations are drawn from the ground-state probability distribution. 
A central question is therefore: does the configuration support of many-body ground states admit a compact representation when configurations are drawn from its probability distribution? If the effective support of the wavefunction grows exponentially with system size, sampling-based subspace constructions may face intrinsic scaling limitations independent of state preparation or measurement inefficiencies.

In this work, we investigate this question by analyzing SQD constructed directly from computational-basis configurations of the exact ground state for representative lattice Hamiltonians. By assuming access to the true ground-state wavefunction, we eliminate problems associated with imperfect reference state preparation and isolate the intrinsic configuration-space structure of the ground-state wavefunction itself. This allows us to determine the minimal number of configurations required to reproduce the ground-state energy within a given accuracy threshold.

We perform this analysis for one- and two-dimensional lattices of the Heisenberg and Hubbard models, two minimal but nontrivial problems often used for benchmarking and scaling analysis purposes in quantum computing, and compare two complementary subspace constructions: deterministic inclusion of configurations in decreasing probability order and stochastic sampling according to the ground-state probability distribution. Across all studied cases, we observe that the number of configurations required to achieve a fixed energy fidelity grows exponentially with system size. Notably, this scaling persists even under idealized, probability-based configuration inclusion, demonstrating that the exponential growth does not originate from sampling inefficiencies but from intrinsic delocalization of the many-body wavefunction in the computational basis.

%These results indicate that SQD effectively probes the configuration-space entropy of the ground state and that the computational-basis support required to represent it within a given accuracy grows exponentially for the lattice models considered here. This shows that sampling-only subspace methods cannot yield polynomial classical post-processing for these models. Our findings suggest that scalable subspace quantum diagonalization for such systems requires basis constructions that extend beyond raw configuration sampling.

The results indicate that sampling-only subspace methods cannot yield polynomial classical post-processing. Our findings suggest that scalable subspace quantum diagonalization for the lattice models considered here requires basis constructions that extend beyond raw configuration sampling.

The remainder of the paper is organized as follows. In \Cref{sec:SQD}, we introduce the sample-based quantum diagonalization algorithm. \Cref{sec:method} describes the lattice Hamiltonians considered in this work and the numerical procedures used to construct SQD subspaces from the exact ground state wavefunction. In \Cref{sec:Results}, we present scaling results for the configuration support required to reproduce ground-state energies of the considered Hamiltonians and analyze their origin. Finally, \Cref{sec:Discussion} discusses the implications of our findings for the scalability of SQD on lattices of the Heisenberg or Hubbard models. 

%------------------------------------------------------------------------
\section{\label{sec:SQD}Sample-Based Quantum Diagonalization}
%------------------------------------------------------------------------
Sample-based quantum diagonalization (SQD) is a hybrid quantum–classical approach for approximating low-energy eigenstates of many-body Hamiltonians by projecting the Hamiltonian onto a configuration subspace generated from measurements of a quantum reference state. Instead of diagonalizing the Hamiltonian in the exponentially large Hilbert space, SQD constructs a reduced subspace that captures the dominant configurations of a reference state and performs classical diagonalization within this subspace~\cite{SQD_original,walkup2026scalingsamplebasedquantumdiagonalization}.

Let $H$ be a Hamiltonian acting on a Hilbert space of dimension $D$. Given a set of basis states $\{\phi_i\}_{i=1}^k$, we define the configuration subspace
\begin{equation}
\mathcal{S}_k = \mathrm{span}\{\phi_1,\phi_2,\dots,\phi_k\}.
\end{equation}
The Hamiltonian projected onto this subspace is
\begin{equation}
H^{(k)}_{ij} = \langle \phi_i | H | \phi_j \rangle .
\end{equation}
Diagonalizing the resulting $k\times k$ matrix yields approximate eigenvalues and eigenvectors of the complete Hamiltonian, with the quality of the approximation determined by how well the chosen subspace captures the relevant components of the true eigenstates. In the limit $k\rightarrow D$, the algorithm performs exact diagonalization in the full Hilbert space, inheriting the limitations discussed prior.

In SQD, the basis states $\{\phi_i\}$ are obtained by sampling computational-basis configurations from a reference quantum state $|\psi_{\mathrm{ref}}\rangle$. Measurements in the computational basis produce configurations $|i\rangle$ with probability
\begin{equation}
p_i = |\langle i | \psi_{\mathrm{ref}} \rangle|^2 .
\end{equation}
Each observed configuration defines a candidate basis vector, and the resulting set of sampled configurations spans the SQD subspace. Repeated sampling, therefore, builds a configuration-space basis that reflects the probability distribution of the reference state.

In practical implementations, the SQD procedure consists of four steps: (1) preparation of a reference state on a quantum device, (2) repeated measurements in the computational basis to generate configurations, (3) construction of the projected Hamiltonian within the resulting configuration subspace, and (4) classical diagonalization of the projected Hamiltonian matrix to obtain approximate eigenvalues.

The construction of the reference state is itself a nontrivial task. It must have significant overlap with the relevant eigenstates and be efficiently preparable on quantum hardware. They may be derived from physical intuition, like Slater determinant~\cite{PhysRev.34.1293,szabo2012modern}, N\'eel state~\cite{auerbach2012interacting}, or Hartree-Fock~\cite{hartree1928wave,fock1930naherungsmethode}, which could be further improved using variational optimization, e.g., by Variational Quantum Eigensolvers and its variants~\cite{peruzzo2014variational,grimsley2019adaptive,PhysRevApplied.19.024047,10.3389/fphy.2022.907160,Jattana_2024,Gaberle_2025}. While the subspace, and therefore, the quality of the solution, is determined by the reference state, there is no generic procedure to prepare a state that guarantees to span the effective subspace without prior knowledge about the problem.

In this work, we consider a simplified setting designed to isolate intrinsic limitations of configuration-space sampling. We skip step (1) and assume access to the exact ground state of the Hamiltonian to be used as reference state, constructing the SQD subspace solely from computational-basis configurations taken from this state. By removing inefficiencies and errors associated with state preparation and measurement, this setting allows us to directly probe how the structure of the ground-state wavefunction determines the configuration support required for accurate subspace diagonalization in an idealized setting.

%------------------------------------------------------------------------
%\subsection{Effects of Sampling State on SQD}%
%\label{sec:Effects of Sampling State on SQD}
%------------------------------------------------------------------------
% What is important? How should the state to sample from look like? What is good, what is bad?
% How to retrieve this state? 
% What happens when taking the true ground state as the sampling state? 
% What insight does this have? Why is it good for the purpose of this paper? 
% How does this generalize?

%------------------------------------------------------------------------
\section{\label{sec:method}Methods}
%------------------------------------------------------------------------
%We compare two scenarios, both taking the ground-state as reference state for a subsequent SQD. In the first, configurations are included deterministically in decreasing probability order, providing an upper bound on the optimal subspace obtainable from this basis. Although this setting is not representative of realistic implementations, it offers a simplified and idealized setting that isolates key features of SQD and facilitates their analysis. In the second, configurations are sampled according to their ground-state probabilities, mimicking measurement-based SQD. This latter setup more closely reflects real-world conditions and thus provides a more realistic assessment of practical performance.

We investigate the configuration-space structure of ground states for representative interacting lattice Hamiltonians. Specifically, we consider spin systems described by the Heisenberg Hamiltonian~\cite{Heisenberg1928} and fermionic systems described by the Hubbard Hamiltonian~\cite{hubbard1963} on one- and two-dimensional lattices. These models provide prototypical settings for strongly correlated quantum systems, remain computationally challenging beyond limited system sizes, and are, therefore, of direct relevance for assessing the potential of SQD methods.

\subsection{Model Hamiltonians}

The Heisenberg Hamiltonian is given by
\begin{equation}
H_{\mathrm{Heis}} = J \sum_{\langle i,j \rangle}
\mathbf{S}_i \cdot \mathbf{S}_j ,
\end{equation}
where $\mathbf{S}_i$ are spin-$\frac{1}{2}$ operators and the sum runs over nearest-neighbor lattice sites. Throughout this work we consider the antiferromagnetic case $J=1$.

For fermionic systems, we consider the Hubbard Hamiltonian
\begin{equation}
H_{\mathrm{Hub}} =
-t \sum_{\langle i,j\rangle,\sigma}
\left(
c^\dagger_{i\sigma} c_{j\sigma} + \mathrm{h.c.}
\right)
+ U \sum_i n_{i\uparrow} n_{i\downarrow},
\end{equation}
where $c^\dagger_{i\sigma}$ and $c_{i\sigma}$ are fermionic creation and annihilation operators, $n_{i\sigma} = c^\dagger_{i\sigma} c_{i\sigma}$ being the spin-density operator for spin $\sigma$ on site $i$, and $U$ denotes the on-site interaction strength. We focus on the intermediate coupling regime $U=2t$ in this study, since it exhibits nontrivial correlations while avoiding strong-/weak-coupling limits, providing a representative and challenging setting.

All Hamiltonians are represented in the computational basis obtained from standard spin-$\frac{1}{2}$ mappings or fermionic occupation-number encodings using Jordan-Wigner transformations~\cite{Jordan1928,generalized_jordan-wigner}. For two-dimensional lattices, sites are ordered along a snake-like path and a spin-up/-down clustering for the Hubbard model, as used in Reference~\cite{Gaberle_2025}.

Lattice sizes range from 6 to 20 spins for 1-D Heisenberg and up to 10 sites for 1-D Hubbard Hamiltonians. For two-dimensional lattices, we consider sizes of $2 \times 2$ up to $4 \times 4$ sites. Therefore, we simulate a maximum of 20 qubits in this study.

\subsection{Ground-State Preparation}

To isolate the intrinsic configuration-space properties of the ground-state wavefunction, we assume access to the exact ground state of each Hamiltonian. Ground states are obtained through classical exact diagonalization of the complete Hamiltonian matrix in the computational basis~\cite{Lanczos,pylanczos}.

This allows direct access to the amplitudes
\begin{equation}
|\psi_0\rangle = \sum_i c_i |i\rangle
\end{equation}
and the corresponding probability distribution
\begin{equation}
p_i = |c_i|^2 .
\end{equation}

Using the exact ground state removes difficulties associated with imperfect state preparation or measurement noise and enables a controlled study of the configuration support required for SQD.

\subsection{Construction of SQD Subspaces}

\begin{figure*}[!ht]
    \centering
    \input{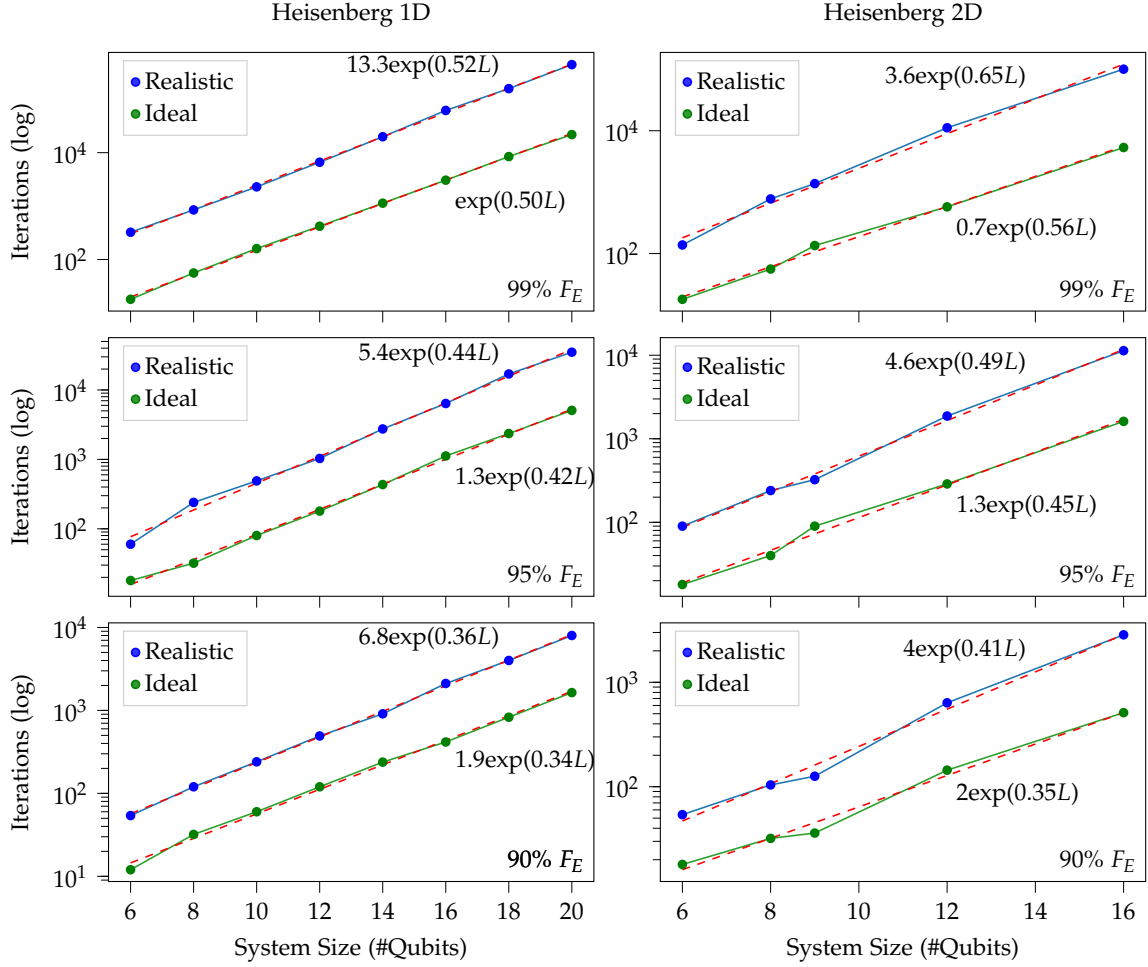}
    \caption{Required configuration samples to span the effective subspace grow exponentially with system size $L$ even under ideal ordering and different $F_E$ thresholds, indicating intrinsic configuration-space complexity of the ground-state for different Heisenberg Hamiltonians. Solid lines correspond to the observed configuration samples necessary, while the dashed line indicates the fitted exponential function written next to it.}
    \label{fig:heis_k-vs-systemsize}
\end{figure*}

We construct SQD subspaces from computational-basis configurations drawn from the ground-state distribution.

Two complementary configuration inclusion strategies are considered:

\emph{Probability-ordered inclusion.}  
Configurations are included deterministically in decreasing order of probability $p_i$. This construction provides a baseline for the optimal configuration compression achievable within the computational basis representation of the ground state for measurement-based SQD. The number of unique configurations equals the number of iterations, $m=k$.

\emph{Sampling-based inclusion.}  
Configurations are sampled according to the probability distribution $p_i$, emulating measurement outcomes obtained from repeated measurements of the reference state on a quantum device. Each newly observed configuration is added to the SQD basis. The number of iterations $m$ can exceed the number of unique configurations sampled, $m\geq k$.

For either strategy, we increase the number of subspace samples in increments proportional to system size. For systems exceeding 16 qubits, however, the increment is fixed at 1000 samples.

%For a given set of configurations $\{\phi_i\}_{i=1}^k$, the Hamiltonian projected onto the resulting subspace is
%
%\begin{equation}
%H^{(k)}_{ij} = \langle \phi_i | H | \phi_j \rangle .
%\end{equation}
%
%Diagonalization of the $k \times k$ projected Hamiltonian yields approximate eigenvalues and eigenvectors.
\newpage
\subsection{Energy Fidelity and Scaling Analysis}

To quantify the accuracy of the SQD approximation, we compare the lowest eigenvalue obtained from the projected Hamiltonian with the exact ground-state energy $E_0$.

We define the energy fidelity as
\begin{equation}
F_E = 1 - \frac{|E_0^k - E_0|}{|E_0|},
\end{equation}
where $E_0^k$ is the lowest eigenvalue of the projected Hamiltonian.

For each system, we determine the minimal number of configuration inclusion strategy iterations $m$ required to reach fixed $F_E$ thresholds of 90\%, 95\%, and 99\%.

The scaling of $m$ with system size $L$ is then analyzed by fitting the numerical data to a function. This procedure allows us to quantify how the configuration support required for accurate subspace diagonalization grows with system size, as well as how efficient measurement-based SQD can accurately span the effective subspace.

\begin{figure*}[t]
    \centering
    \input{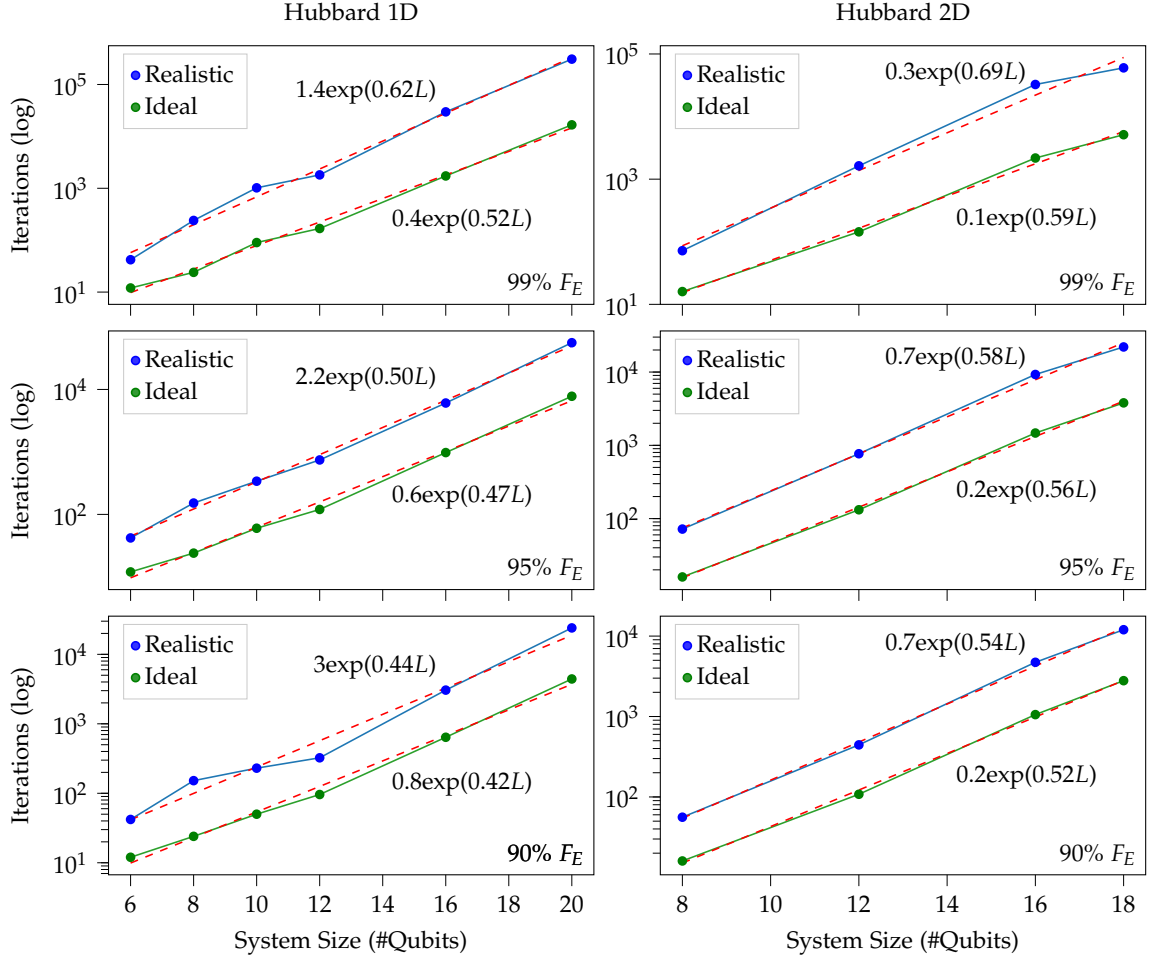}
    \caption{Required configuration samples to span the effective subspace grow exponentially with system size $L$ even under ideal ordering and different energy fidelity thresholds, indicating intrinsic configuration-space complexity of the ground-state for different Hubbard Hamiltonians. Solid lines correspond to the observed configuration samples necessary, while the dashed line indicates the fitted exponential function written next to it.}
    \label{fig:hub_k-vs-systemsize}
\end{figure*}

%------------------------------------------------------------------------
\section{\label{sec:Results}Results}%
%------------------------------------------------------------------------
% Explain the data generation. Why did we use the true ground state? 
% What is the implication from using the true ground state as our sample state for SQD? Give mathematical reasoning as to why this is useful, what are the drawbacks and the implications from it. 
% Why decided for this? What is the key insight we can expect from this?
% Show the results, i.e., plots
% Explain the implication, like exponential increase in measurement overhead
% Give a rough estimate of problem sizes applicable for different accuracies
% What does this mean for the applicability of SQD to lattice models

We compare two scenarios, both taking the ground state as the reference state for a subsequent SQD. In the first, configurations are included deterministically in decreasing probability order, providing an upper bound on the optimal subspace obtainable from this basis. Although this setting is not representative of realistic implementations, it offers a simplified and idealized setting that isolates key features of SQD and facilitates their analysis. In the second, configurations are sampled according to their ground-state probabilities, mimicking measurement-based SQD. This latter setup more closely reflects real-world conditions and thus provides a more realistic assessment of practical performance.

%------------------------------------------------------------------------
\subsection{Exponential Scaling of Required Configuration Support}
%------------------------------------------------------------------------

We first analyze the number of computational-basis configuration samples required to achieve a fixed target accuracy in the ground-state energy using SQD. For each system size, we determine the minimal number $m$ to span the subspace $S_k$ necessary to reach 99\%, 95\%, and 90\% energy fidelity relative to the exact ground-state energy.

\Cref{fig:heis_k-vs-systemsize} shows the scaling of the required subspace expansion iterations $m$ with system size $L$ for both probability-ordered (ideal) inclusion and stochastic sampling (realistic) for Heisenberg model Hamiltonians. The ideal ordering,n therefore, provides the optimal compression achievable within the computational basis representation of the exact ground state. Across all systems and fidelity thresholds, we observe clear exponential scaling:
\begin{equation}
    m(L) \sim \exp(\alpha L), 
\end{equation}
with comparable exponents for both sampled and ideally ordered bases. Latter demonstrates that the growth of $m$ is not induced by statistical sampling inefficiency but instead reflects intrinsic configuration-space complexity of the ground-state. 

Similarly, the scaling of $m$ to the system size in numbers of qubits for the Hubbard model can be observed in \Cref{fig:hub_k-vs-systemsize}. In particular, both one- and two-dimensional systems of the Heisenberg and Hubbard models exhibit similar scaling behavior, indicating that this phenomenon is not restricted to a specific geometry or particle statistics. 

Based solely on the coefficients of the fitted exponential functions, the Hubbard model lattices appear harder to solve than the Heisenberg model lattices of the same size. The exponents are also consistently increasing for higher $F_E$ thresholds, whereby the scaling of the coefficients is not linear. Lastly, for each model, the 1-D case exhibits smaller exponents in the exponential scaling compared to 2-D Hamiltonians of similar size throughout all experiments. 

\begin{figure}[ht]
    \centering
    % This file was created with matplot2tikz v0.5.3.
\begin{tikzpicture}[scale=1, transform shape]

\definecolor{darkgray176}{RGB}{176,176,176}
\definecolor{darkorange25512714}{RGB}{255,127,14}
\definecolor{lightgray204}{RGB}{204,204,204}
\definecolor{steelblue31119180}{RGB}{31,119,180}

\begin{groupplot}[group style={
    group size=2 by 2,
    horizontal sep=10pt,
    vertical sep=1.5cm
}]
\nextgroupplot[
height=4.5cm,
width=4.5cm,
legend cell align={left},
legend style={
  fill opacity=0.8,
  draw opacity=1,
  text opacity=1,
  at={(0.03,0.97)},
  anchor=north west,
  draw=lightgray204
},
log basis y={10},
tick align=outside,
tick pos=left,
title={Hubbard 1D},
x grid style={darkgray176},
xmin=5.3, xmax=20.7,
xtick style={color=black},
xtick={10,15,20},
xticklabels={
  \(\displaystyle {10}\),
  \(\displaystyle {15}\),
  \(\displaystyle {20}\),
},
y grid style={darkgray176},
ylabel={Subspace Size (log)},
ymin=5.40428766469088, ymax=44789.2850514899,
ymode=log,
ytick style={color=black}
]
\addplot [semithick, steelblue31119180]
table {%
6 9
8 41
10 86
12 179
16 1973
20 20231
};
\addlegendentry{$k$}
\addplot [semithick, darkorange25512714]
table {%
6 8.14421387954949
8 15.6704890327681
10 56.8153021506125
12 108.468059897005
16 809.19724767456
20 5749.41610148327
};
\addlegendentry{$N_{eff}$}

\nextgroupplot[
height=4.5cm,
width=4.5cm,
legend cell align={left},
legend style={
  fill opacity=0.8,
  draw opacity=1,
  text opacity=1,
  at={(0.03,0.97)},
  anchor=north west,
  draw=lightgray204
},
log basis y={10},
scaled y ticks=manual:{}{\pgfmathparse{#1}},
tick align=outside,
tick pos=left,
title={Hubbard 2D},
x grid style={darkgray176},
xmin=7.5, xmax=18.5,
xtick style={color=black},
xtick={10,14,18},
xticklabels={
  \(\displaystyle {10}\),
  \(\displaystyle {14}\),
  \(\displaystyle {18}\),
},
y grid style={darkgray176},
ymin=5.40428766469088, ymax=44789.2850514899,
ymode=log,
ytick style={color=black},
yticklabels={}
]
\addplot [semithick, steelblue31119180]
table {%
8 16
12 151
16 2362
18 5403
};
\addlegendentry{$k$}
\addplot [semithick, darkorange25512714]
table {%
8 15.3197113699419
12 113.577552054111
16 1100.25671838656
18 2917.81798610757
};
\addlegendentry{$N_{eff}$}

\nextgroupplot[
height=4.5cm,
width=4.5cm,
legend cell align={left},
legend style={
  fill opacity=0.8,
  draw opacity=1,
  text opacity=1,
  at={(0.03,0.97)},
  anchor=north west,
  draw=lightgray204
},
log basis y={10},
tick align=outside,
tick pos=left,
title={Heisenberg 1D},
x grid style={darkgray176},
xlabel={System Size (\#Qubits)},
xmin=5.3, xmax=20.7,
xtick style={color=black},
xtick={10,15,20},
xticklabels={
  \(\displaystyle {10}\),
  \(\displaystyle {15}\),
  \(\displaystyle {20}\),
},
y grid style={darkgray176},
ylabel={Subspace Size (log)},
ymin=5.40428766469088, ymax=44789.2850514899,
ymode=log,
ytick style={color=black}
]
\addplot [semithick, steelblue31119180]
table {%
6 18
8 54
10 160
12 472
14 1327
16 3877
18 10704
20 29721
};
\addlegendentry{$k$}
\addplot [semithick, darkorange25512714]
table {%
6 11.1677071209291
8 26.2030529106278
10 61.54125306387
12 138.588393372538
14 313.778101622185
16 709.45651815247
18 1642.70541635892
20 3644.0933146471
};
\addlegendentry{$N_{eff}$}

\nextgroupplot[
height=4.5cm,
width=4.5cm,
legend cell align={left},
legend style={
  fill opacity=0.8,
  draw opacity=1,
  text opacity=1,
  at={(0.03,0.97)},
  anchor=north west,
  draw=lightgray204
},
log basis y={10},
scaled y ticks=manual:{}{\pgfmathparse{#1}},
tick align=outside,
tick pos=left,
title={Heisenberg 2D},
x grid style={darkgray176},
xlabel={System Size (\#Qubits)},
xmin=5.5, xmax=16.5,
xtick style={color=black},
xtick={8,12,16},
xticklabels={
  \(\displaystyle {8}\),
  \(\displaystyle {12}\),
  \(\displaystyle {16}\),
},
y grid style={darkgray176},
ymin=5.40428766469088, ymax=44789.2850514899,
ymode=log,
ytick style={color=black},
yticklabels={}
]
\addplot [semithick, steelblue31119180]
table {%
6 17
8 56
9 131
12 642
16 6560
};
\addlegendentry{$k$}
\addplot [semithick, darkorange25512714]
table {%
6 11.1009361395447
8 25.7229965218292
9 45.573870488639
12 148.753684654791
16 811.458240138056
};
\addlegendentry{$N_{eff}$}
\end{groupplot}

\end{tikzpicture}
    \caption{Effective number of configurations $N_{eff}$ grows exponentially with system size. The required subspace size $k$ therefore grows exponentially with system size too, since $k \gtrsim N_{eff}$. }
    \label{fig:shannon-entropy}
\end{figure}
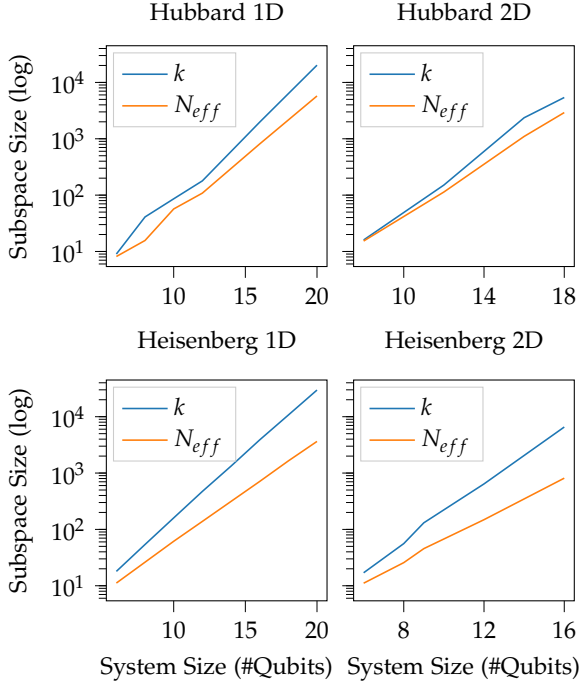
%------------------------------------------------------------------------
\subsection{Origin of Configuration-Space Complexity}
%------------------------------------------------------------------------

This scaling can be interpreted quantitatively through the configuration-space entropy of the ground-state wavefunction. Writing the exact ground state as
\begin{equation}
|\psi_0\rangle = \sum_i c_i |i\rangle,
\end{equation}
with probabilities $p_i = |c_i|^2$, the spread of the wavefunction over basis configurations can be quantified by the von Neumann entropy on eigenvectors~\cite{vonNeumann}, generalizing to the Shannon entropy of the eigenvalues~\cite{shannon}
\begin{equation}
S = -\sum_i p_i \log p_i .
\end{equation}
The quantity $N_{\mathrm{eff}} = e^{S}$ defines the effective number of configurations that significantly contribute to the wavefunction, generalizing the notion of support size to non-uniform probability distributions.

For a sampling-based construction of the subspace, configurations are drawn according to $p_i$. In order for the resulting subspace to faithfully represent the low-energy eigenspace, it must include a representative fraction of this effective support. Consequently, the number of configurations required to achieve high-fidelity reconstruction $N_{\mathrm{eff}}$ must satisfy approximately
\begin{equation}
k \gtrsim N_{\mathrm{eff}} = e^{S},
\end{equation}
while the subspace size $k$ has to span the effective configurations at least. This relation can be observed in \Cref{fig:shannon-entropy} where we plotted $N_{\mathrm{eff}}$ and $k$ against the respective model's system size. For interacting lattice systems, the entropy of the ground-state distribution grows extensively with system size, $S \propto L$, implying an exponential growth of the effective configuration support. The exponential scaling of the required subspace size observed in \Cref{fig:heis_k-vs-systemsize,fig:hub_k-vs-systemsize} therefore follows naturally from the delocalized structure of the many-body wavefunction in the computational basis.

%------------------------------------------------------------------------
\subsection{Absence of a Sparse Dominant Configuration Manifold}
%------------------------------------------------------------------------

% Plot energy fidelity vs cumulative probability mass of the ground-state (ideal case)
\begin{figure*}[htbp]
    \centering
    \input{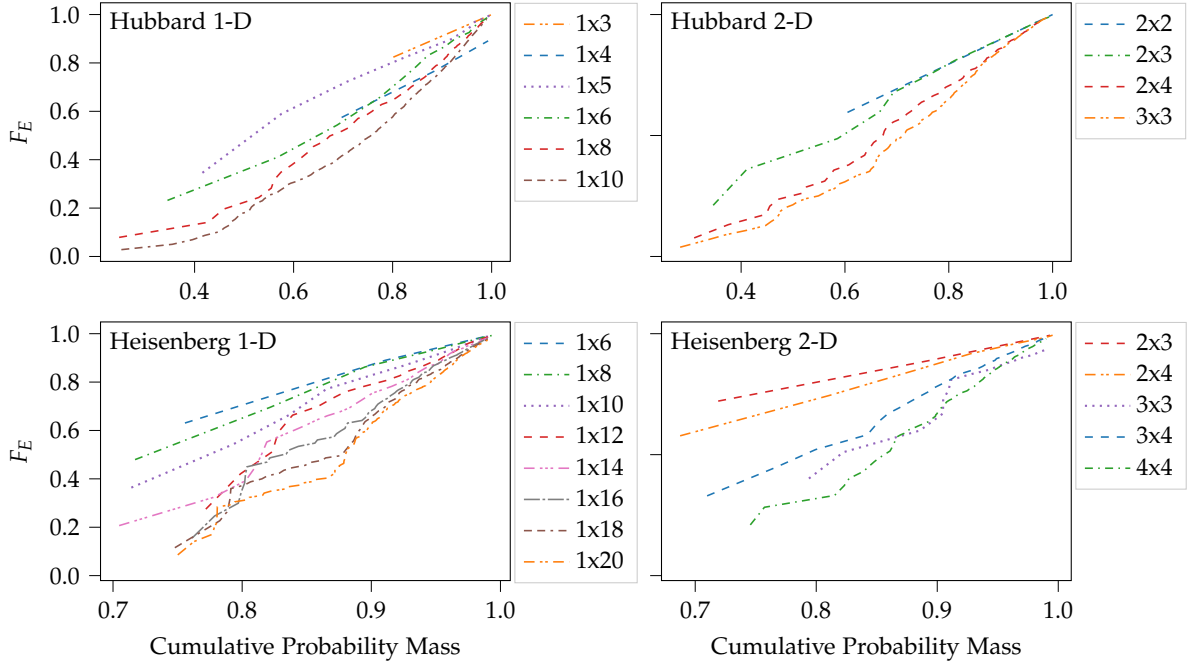}
    \caption{$F_E$ plotted against the cumulative probability mass captured in the subspace configurations for different lattice sizes, written as a combination of height and width. Accurate reconstruction of the ground-state energy requires the inclusion of a large fraction of the ground-state probability mass, demonstrating the absence of a sparse dominant configuration manifold. Larger lattice sizes exhibit a larger $N_{\mathrm{eff}}$, therefore each added configuration has a lesser impact overall, explaining the lower $F_E$ at the beginning.}
    \label{fig:energy-vs-cumprob}
\end{figure*}

The ordered-configuration construction used in our analysis approximates this optimal support directly by deterministically including the most probable configurations. The close agreement between ordered and sampled scaling, therefore, indicates that SQD performance is fundamentally limited by the intrinsic configuration-space entropy of the ground state rather than by sampling inefficiencies.

This can be readily observed in \Cref{fig:energy-vs-cumprob}, which shows the relationship between cumulative probability mass, referring to the sum of probabilities each configuration of the subspace carries in the ground state, and energy fidelity. For each system, we compute the cumulative probability mass captured by the included configurations and track the corresponding energy fidelity obtained from diagonalization in the restricted subspace.

%We observe that the high-fidelity reconstruction requires inclusion of a large fraction of the total probability mass, typically exceeding 90\% and often approaching near-complete coverage. No evidence is found for a sparse dominant core of configurations that alone determine the ground-state energy.

No small subset of high-probability configurations suffices to accurately reproduce the ground-state energy. Instead, high energy fidelity is achieved only after incorporating a broad distribution of configurations, spanning the dominant support of the wavefunction, whose total number grows exponentially with system size.

%This behavior demonstrates that the energy contribution are broadly distributed across the configuration space. In other words, the ground-state wavefunction does not admit a small, high-weight subset of configurations sufficient for variational completeness. Instead, accurate reconstruction requires incorporation of a wide configuration manifold whose size grows exponentially with system size.

%This demonstrates that the ground-state wavefunction is delocalized in the computational basis and lacks a sparse dominant support capable of encoding the low-energy eigenspace.

% Version 1
While the delocalized nature of these ground states is well understood from the perspectives of entanglement structure and area laws~\cite{Hastings_2007,RevModPhys.82.277}, as well as from studies of Hilbert-space participation and eigenspace complexity~\cite{HS_participation,Beugeling_2015}, its implications for configuration-space sparsity and sampling-based quantum algorithms remain comparatively unexplored in the scientific literature. Our results provide this, as they explicitly show that the wavefunction lacks a sparse dominant support capable of efficiently encode the low-energy eigenspace.

% Version 2
%This observation is consistent with the well-known fact that ground states of interacting lattice models, such as the Heisenberg and Hubbard Hamiltonians, are coherent superpositions of a large number of configurations, as reflected, for example, in valence-bond descriptions and participation-ratio analyses. However, our results make this structure explicit in the computational basis and demonstrate that the wavefunction lacks a sparse dominant support capable of efficiently encoding the low-energy eigenspace.

%While the delocalized nature of these ground states is well understood from the perspective of entanglement and Hilbert-space participation, its implications for configuration-space sparsity and sampling-based quantum algorithms have received comparatively little attention.

This observation also directly explains the exponential scaling seen in \Cref{fig:heis_k-vs-systemsize,fig:hub_k-vs-systemsize}. As system size increases, the number of configurations required to accumulate the necessary probability mass grows exponentially.

%------------------------------------------------------------------------
\subsection{Sampling Efficiency}
%------------------------------------------------------------------------

% Plot sampling efficiency, i.e., number of measurements vs unique basis states
\begin{figure*}
    \centering
    \input{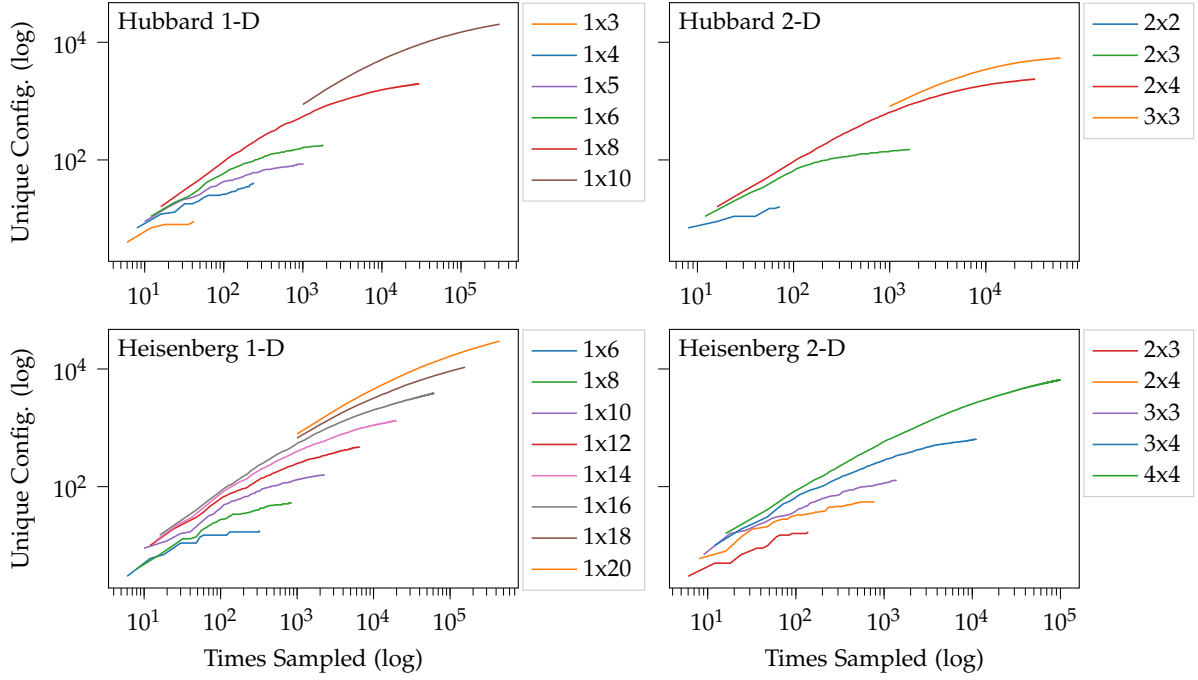}
    \caption{Sampling saturates dominant configurations, but to achieve high accuracy, the inclusion of less dominant configurations is necessary, as observed with the flattening of the curves at the end, indicating that the exponential scaling originates primarily from intrinsic wavefunction structure rather than measurement inefficiency.}
    \label{fig:sampling_efficiency}
\end{figure*}

To distinguish intrinsic wavefunction structure from possible measurement inefficiencies, \Cref{fig:sampling_efficiency} shows the number of unique configurations discovered as a function of measurement steps. For all systems, sampling rapidly saturates the dominant configurations, indicating that high-probability states are efficiently identified.

%For all systems, the number of unique basis states grows rapidly at small measurement counts but subsequently saturates as dominant configurations are repeatedly sampled. This saturation indicates that the measurement-based sampling efficiently captures the high-probability configurations early. 

The exponential scaling of the required subspace size, therefore, cannot be attributed to poor discoverability of important configurations. Rather, the dominant configurations are readily sampled, but they are insufficient for high-fidelity energy reconstruction. Additional configurations, each carrying a smaller individual probability, are nevertheless essential for completeness. 

Our results indicate that sampling-only subspace quantum diagonalization effectively probes the configuration-space entropy of the many-body wavefunction. The exponential growth of the required configuration support reflects intrinsic delocalization rather than limitations of the algorithm itself.

%A notable feature of our results is the qualitative similarity across one- and two-dimensional systems, spin systems described by the Heisenberg model, and fermionic systems described by the Hubbard model. Despite differing microscopic degrees of freedom and lattice geometries, all studied systems exhibit exponential scaling of required configuration support. This suggests that the absence of polynomial configuration compression is not model-specific but rather reflects a general property of interacting lattice ground-states in the computational basis.
%
%We find that the number of configurations required to achieve variational accuracy grows exponentially with system size in both scenarios and either lattice model. The persistence of exponential scaling even under ideal ordering demonstrates that this behavior originates from intrinsic delocalization of the ground-state wavefunction in the computational basis rather than from statistical sampling inefficiencies.

%------------------------------------------------------------------------
\section{\label{sec:Discussion}Discussion}%
%------------------------------------------------------------------------
% What is the key takeaway from the results prior? One sentence starter
In this work, we have investigated the configuration-space structure underlying sample-based quantum diagonalization by constructing subspaces directly from computational-basis configurations of exact ground states. By eliminating reference state preparation and real-world measurement inefficiencies, our analysis isolates the intrinsic properties of the many-body wavefunction that govern the effectiveness of SQD methods.

Our central finding is that the number of configurations required to achieve high accuracy in the ground-state energy grows exponentially with system size for both Heisenberg and Hubbard model Hamiltonians. Importantly, this exponential scaling persists even when configurations are included in optimal probability-ordered fashion. This demonstrates that the observed growth is not a consequence of statistical sampling inefficiencies, but instead reflects the intrinsic property of the ground-state wavefunction, being its delocalization in the computational basis and the absence of a sparsely supported configuration manifold capable of capturing the low-energy eigenspace. 

These results indicate that the low-energy eigenspace of the studied lattice Hamiltonians cannot be represented within a polynomially bounded configuration subspace derived from the ground-state distribution. Consequently, sampling-only SQD in the computational basis faces an intrinsic scaling limitation independent of its reference state preparation, which itself remains a matter of active research~\cite{LaRose2019,miura2026activesamplingsamplebasedquantum,cantori2025adaptivebasissamplebasedneuraldiagonalization}, or measurement noise.

% What does this mean for the application of this method?
%Our findings suggest that scalable SQD for these models requires basis constructions that extend beyond the raw configuration support of the ground state, for example, through operator-generated states, symmetry-adapted bases, or measurement in rotated bases.

This behavior can be understood in terms of the statistical structure of the wavefunction. Accurate reconstruction of the ground-state energy requires the inclusion of configurations that collectively carry a substantial fraction of the total probability mass. As the system size increases, this probability mass is distributed over an exponentially growing number of configurations, implying that no polynomially bounded subset of computational-basis states suffices for completeness. In this sense, sampling-based subspace constructions probe an effective configuration-space entropy of the ground state, which sets a fundamental limit on the achievable compression within this basis. 

These results have direct implications for the scalability of subspace quantum diagonalization methods for the problems we tested. In its simplest form, where the subspace is constructed solely from configurations obtained via measurements in the computational basis, SQD cannot yield a polynomially scaling classical post-processing cost for the lattice models considered here. This limitation is intrinsic and persists even under idealized conditions, indicating that improvements in the sampling strategy alone are insufficient to overcome it. 

At the same time, our findings do not preclude the broader applicability of subspace quantum diagonalization methods. Rather, they identify a specific limitation of sampling-only constructions and thereby clarify the requirements for scalable approaches. In particular, our results suggest that successful subspace methods must incorporate additional structure beyond raw configuration sampling, such as operator-generated basis states~\cite{partitioned_QSE,SKQD,SKQD_Heisenberg}, symmetry-adapted subspaces~\cite{symmetry_adapted_SQD,Akande_2025}, or measurements in rotated bases~\cite{cantori2025adaptivebasissamplebasedneuraldiagonalization} that more efficiently capture the relevant degrees of freedom.

Finally, the consistency of our observations across one- and two-dimensional systems, as well as across spin and fermionic models, suggests that the absence of polynomial configuration compression is not model-specific but instead reflects a general feature of interacting lattice ground states in the computational basis. Investigating the connection between this behavior and measures of entanglement, computational bases, or configuration-space entropy represents an interesting direction of future work. 

Taken together, our results provide a structural perspective on the limitations of SQD and highlight the importance of basis design, i.e., choosing a representation in which the wavefunction has low effective support, in quantum-classical algorithms for many-body systems.

%---------------------------------------------------------------------------------------
%	 REFERENCES
%---------------------------------------------------------------------------------------

\printbibliography
%---------------------------------------------------------------------------------------

\end{document}